\documentclass[twocolumn, aps, pra, showpacs, superscriptaddress, floatfix, nofootinbib, 10pt]{revtex4-2}
\usepackage{times}

\usepackage{amsmath}
\usepackage{graphicx}
\usepackage{floatrow}
  \newfloatcommand{capbtabbox}{table}[][\FBwidth]
\usepackage{diagbox}
\usepackage{xcolor}
  \definecolor{purple}{RGB}{128,0,128}
\usepackage{braket}
\usepackage{bbold}
\usepackage[hidelinks]{hyperref}

\title{XXX}

\date{}

\usepackage{changes}

\begin{document}

\title{Experimental genuine quantum nonlocality in the triangle network}

\author{Ning-Ning Wang}
\affiliation{Laboratory of Quantum Information, University of Science and Technology of China, Hefei 230026, China}
\affiliation{Anhui Province Key Laboratory of Quantum Network, University of Science and Technology of China, Hefei 230026, China}
\affiliation{CAS Center for Excellence in Quantum Information and Quantum Physics, University of Science and Technology of China, Hefei 230026, China}
\affiliation{Hefei National Laboratory, University of Science and Technology of China, Hefei 230088, China}
\author{Chao Zhang}\email{drzhang.chao@ustc.edu.cn}
\affiliation{Laboratory of Quantum Information, University of Science and Technology of China, Hefei 230026, China}
\affiliation{Anhui Province Key Laboratory of Quantum Network, University of Science and Technology of China, Hefei 230026, China}
\affiliation{CAS Center for Excellence in Quantum Information and Quantum Physics, University of Science and Technology of China, Hefei 230026, China}
\affiliation{Hefei National Laboratory, University of Science and Technology of China, Hefei 230088, China}
\author{Huan Cao}
\affiliation {State Key Laboratory of Optoelectronic Materials and Technologies, School of Electronics and Information Technology, Sun Yat-sen University, Guangzhou 510006, China}
\author{Kai Xu}
\affiliation{Laboratory of Quantum Information, University of Science and Technology of China, Hefei 230026, China}
\affiliation{Anhui Province Key Laboratory of Quantum Network, University of Science and Technology of China, Hefei 230026, China}
\affiliation{CAS Center for Excellence in Quantum Information and Quantum Physics, University of Science and Technology of China, Hefei 230026, China}
\author{Bi-Heng Liu}
\affiliation{Laboratory of Quantum Information, University of Science and Technology of China, Hefei 230026, China}
\affiliation{Anhui Province Key Laboratory of Quantum Network, University of Science and Technology of China, Hefei 230026, China}
\affiliation{CAS Center for Excellence in Quantum Information and Quantum Physics, University of Science and Technology of China, Hefei 230026, China}
\affiliation{Hefei National Laboratory, University of Science and Technology of China, Hefei 230088, China}
\author{Yun-Feng Huang}\email{hyf@ustc.edu.cn}
\affiliation{Laboratory of Quantum Information, University of Science and Technology of China, Hefei 230026, China}
\affiliation{Anhui Province Key Laboratory of Quantum Network, University of Science and Technology of China, Hefei 230026, China}
\affiliation{CAS Center for Excellence in Quantum Information and Quantum Physics, University of Science and Technology of China, Hefei 230026, China}
\affiliation{Hefei National Laboratory, University of Science and Technology of China, Hefei 230088, China}
\author{Chuan-Feng Li}\email{cfli@ustc.edu.cn}
\affiliation{Laboratory of Quantum Information, University of Science and Technology of China, Hefei 230026, China}
\affiliation{Anhui Province Key Laboratory of Quantum Network, University of Science and Technology of China, Hefei 230026, China}
\affiliation{CAS Center for Excellence in Quantum Information and Quantum Physics, University of Science and Technology of China, Hefei 230026, China}
\affiliation{Hefei National Laboratory, University of Science and Technology of China, Hefei 230088, China}
\author{Guang-Can Guo}
\affiliation{Laboratory of Quantum Information, University of Science and Technology of China, Hefei 230026, China}
\affiliation{Anhui Province Key Laboratory of Quantum Network, University of Science and Technology of China, Hefei 230026, China}
\affiliation{CAS Center for Excellence in Quantum Information and Quantum Physics, University of Science and Technology of China, Hefei 230026, China}
\affiliation{Hefei National Laboratory, University of Science and Technology of China, Hefei 230088, China}
\author{Nicolas Gisin}
\affiliation{Group of Applied Physics, University of Geneva, 1211 Geneva 4, Switzerland}
\affiliation{Constructor University, Bremen, Germany}
\author{Tam\'as Kriv\'achy}\email{tamas.krivachy@gmail.com}
\affiliation{ICFO - Institut de Ciencies Fotoniques, The Barcelona Institute of Science and Technology, 08860 Castelldefels (Barcelona), Spain}
\affiliation{Atominstitut, Technische Universität Wien, 1020 Vienna, Austria}
\author{Marc-Olivier Renou}\email{marc-olivier.renou@inria.fr}
\affiliation{CPHT, LIX, CNRS, Inria, École polytechnique, Institut Polytechnique de Paris, Palaiseau, France
}

\begin{abstract}
In the last decade, it was understood that quantum networks involving several independent sources of entanglement which are distributed and measured by several parties allowed for  completely novel forms of nonclassical quantum correlations, when entangled measurements are performed. 
Here, we experimentally obtain quantum correlations in a triangle network structure, and provide solid evidence of its nonlocality. 
Specifically, we first obtain the Elegant distribution proposed in~\cite{Gisin2019} 
by performing a six-photon experiment.
Then, we justify its nonlocality based on machine learning tools to estimate the distance of the experimentally obtained correlation to the local set.
\end{abstract}
\maketitle

\emph{Introduction.---} Bell theorem proved that quantum theory's operational implications are irreconcilable with any Local Hidden Variable (LHV) model, or explanation. More precisely, two distant parties (call them Alice and Bob) measuring an appropriate entangled quantum system can observe fundamentally nonclassical space-like separated correlated events, called nonlocal correlations. 
Bell’s radical conclusion was later confirmed by a series of experiments~\cite{Freedman1972, Aspect1982, loophole1, loophole2, loophole3}, finally recognized by the Nobel committee in 2022.
This milestone theorem had also a profound impact on our understanding of what quantum correlations are or allow for, both for foundational reasons and concrete applications~\cite{DIQKD, DIQRNG}.

\begin{figure}
    \centering
    \includegraphics[width=\textwidth]{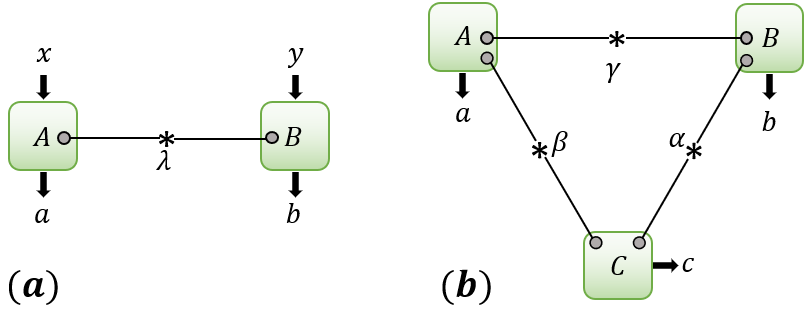}
    \caption{(a) Classical LHV strategy in a standard Bell scenario. The source (star) sample a random variable $\lambda$ and send its value to $A,B$. $A$ outputs $a$, a function of her input $x$ and the value of $\lambda$, $B$ does similarly. 
    With an appropriate bipartite quantum state and local measurements, $A,B$ can obtain correlations with no LHV model explanation.\\
    (b) Classical triangle network LHV strategy . 
    The three sources (star) respectively sample independent random variables $\alpha, \beta, \gamma$ and send their values to $A, B, C$ according to the triangle causal structure. 
    $A$ outputs $a$, a function of the values of $\beta, \gamma$ and $B, C$ do similarly. 
    This allows them to share a probability distribution as in Eq.~\ref{eq:local_model}.
    In quantum triangle network strategies, the sources produce independent bipartite quantum states, allowing $A, B, C$ to share a probability distribution as in Eq.~\ref{eq:quantum_model}.
    The EJM distribution of Eq.~\ref{eq:EJM_Distribution} has a quantum model and is expected to have no classical triangle network LHV strategy explanation.
}
    \label{fig:scenarios}
\end{figure}

More recently, it was understood that beyond the standard Bell scenario in which several parties measure a unique quantum state to establish correlations between them, other more general approaches to nonlocality could be considered. 
In particular, quantum networks, in which several independent sources are distributed to the parties, were shown to display a new form of nonlocality~\cite{NNrev, Branciard2010, Branciard2012, Tavakoli2014, Chaves2016, Rosset2016}. 
More precisely, there exists network nonlocal probability distributions, that are distributions obtained by local measurements on several independent quantum sources which have no explanation in terms of classical network LHV strategies. 
This manifests even without inputs, in particular in the triangle network of Fig.~\ref{fig:scenarios}~\cite{Marco2019, TCCM, TCCM2, TCCM3}. 

The concept of quantum networks allowed new developments in quantum foundations such as several generalizations of the Bell theorem to exclude other alternatives to quantum theory, beyond LHV models~\cite{Mirjam, Real_Expt, Real2021, Xavier2021, Xavier2021pra, GMNL_Experiment}. 
It also enabled new key applications to quantum correlations~\cite{EntangledMeasurements, Pavel2023}, e.g. providing strong arguments in favor of the certifiability (self testing) of all pure quantum states~\cite{Supic2023}, a long standing open question.

While well understood algorithms exist to analyse what probability distributions admit a LHV in the standard Bell scenario, characterising correlations in networks is much harder, the problem being not convex. 
Hence, most analytical proofs of network nonlocality are only valid in the perfect noiseless, infinite statistics case. 
Some numerical tools exist, such as the inflation method \cite{InflationTechnic}, but their numerical complexity make their practical use difficult \cite{InflationImplementation}.
More recently, machine learning heuristics were proposed, which construct explicit LHV models, and can find an approximately best LHV model to minimize a given objective function, such as an inequality or the distance to a target distribution~\cite{NNpaper, 2024_ineq_Tamas}.
These heuristics are the most efficient approaches to understand whether a generic distribution has a local explanation or not, due to their performance in network scenarios. 
They have led to conjectures of nonlocality that have since been proven~\cite{pozas2023proofs}. 
They are in practice the only tool to study noise robustness of distributions and nonlocality subject to realistic, experimental environments~\cite{singlephotontriangle}.

Few experimental proofs of quantum network nonlocality were performed, in the entanglement swapping scenario~\cite{saunders2017experiment,carvacho2017experiment} (with several loopholes closed in~\cite{sun2019experiment}), the star networks~\cite{poderini2020experiment, Chao2021experiment} and triangle networks~\cite{suprano2022experiment}.
However these are all closely related to standard violation of Bell theorem in a scenario involving a single source and can be realized without entangled measurements. E.g.,~\cite{ Polino2022experiment} implements the Fritz distribution, which can be viewed as a standard Bell test embedded in the triangle network, in which only one source needs to be entangled and the other two sources can be classically correlated.
In the bilocal scenario, several attempt were performed to solve this problem. 
In particular, the correlations generated by the Elegant Joint Measurements (EJMs) (see Eq.~\ref{eq:EJM_Measurement} which was studied in~\cite{bilocalEJM, CenXiao2022experiment}) and a new concept of full network nonlocality was proposed~\cite{Fullnetwork} and demonstrated~\cite{Xuemei2023experiment,NingNing2023experiment}.
All these experiments try to obtain genuine network nonlocal correlations, that is nonlocal correlations obtained in a network which cannot be viewed as coming from some embedded standard Bell test in a network~\cite{Marco2019, NNrev}.

Here, we go beyond the bilocal model and study the correlations generated in the triangle network. Different from previous experimental studies of triangle network, we consider entangled measurements at each node. Specifically, all three parties are pairwise connected via a polarisation entangled photon source and at each party we perform an EJM. The generated correlation (see Eq.~\ref{eq:EJM_Distribution}) is believed to be genuine to the triangle network, that is not related to standard Bell nonlocality~\cite{NNrev, GNN}. 
We demonstrate the nonlocality of the observed correlations based on a machine-learning-based heuristic program which calculates the minimal distance between the observed correlation to the local hidden variable models. Our results demonstrate that experimentally observing this new form of nonlocality is possible with current technology.

\begin{figure}
  \centering
  \includegraphics[scale = 0.085]{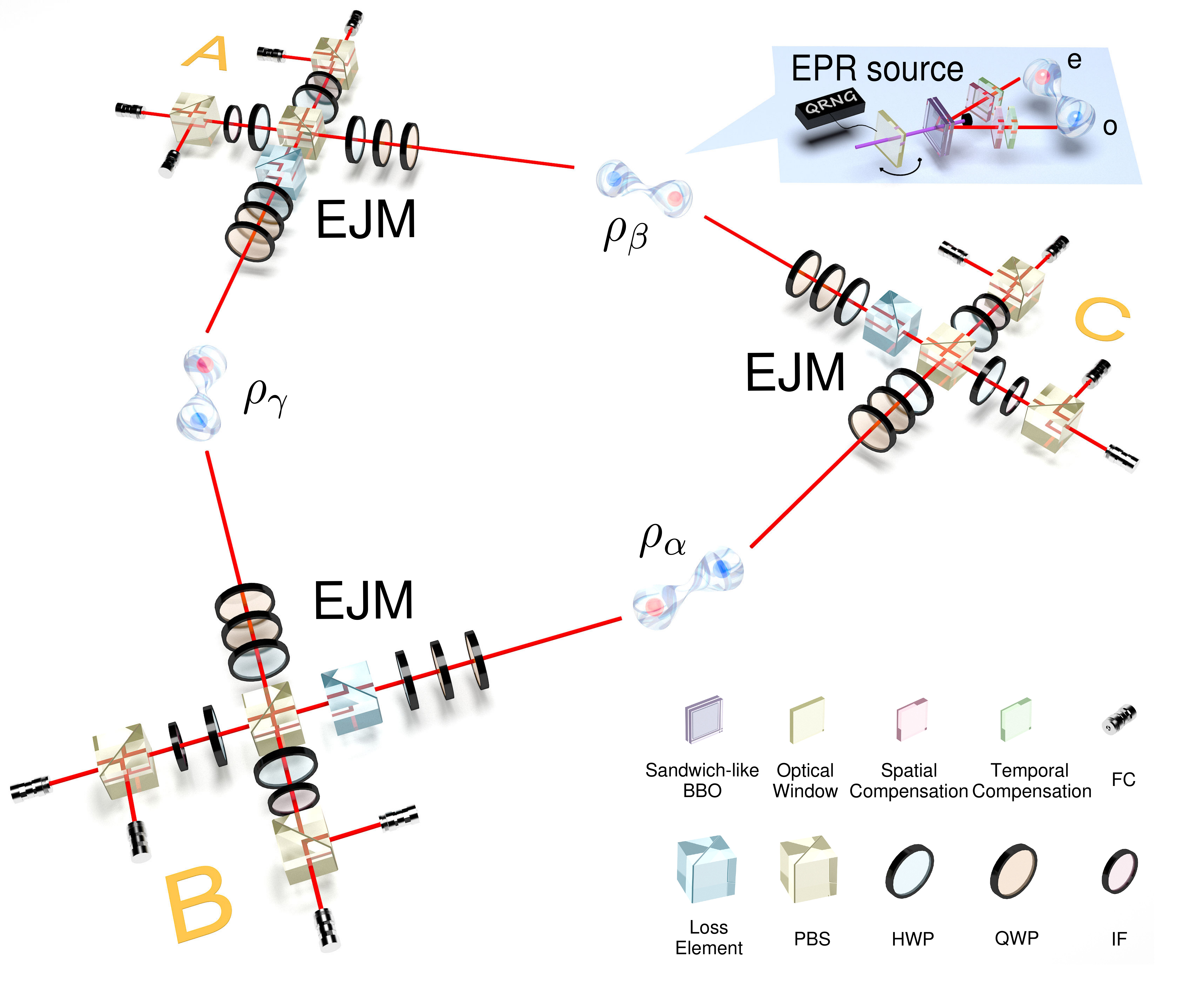}
  \caption{Sketch of the experimental setup. Three EPR sources distribute biphoton singlet state through SPDC process. 
  The EPR source has a sandwich-like BBO-HWP-BBO structure with 390-nm, 80-MHz, 140-fs ultraviolet pulse pumping. An optical window controlled by a QRNG randomizes the 
  phase of the pump pulse before it is incident on the crystal. Each node of the network receives an ordinary photon (red sphere) and an extraordinary photon (blue sphere), which come from two different sources. 
  At each node, the partially entangled projection is achieved through the polarization-dependent loss element and photonic Hong-Ou-Mandel interference, and the EJM is achieved by switching the projection settings. 
  FC fiber coupler, PBS polarization beam splitter, HWP half-wave plate, IF interference filter, QRNG quantum random number generator, BBO beta barium borate. }
  \label{setup}
\end{figure}

\emph{Network nonlocality in the triangle.---} In the triangle network, with classical bipartite sources, the three parties are able to sample from probability distributions of the form
\begin{align}\label{eq:local_model}
    P(a,b,c) = &\int d\tilde\alpha d\tilde\beta d\tilde\gamma P_{\alpha}(\tilde\alpha) P_{\beta}(\tilde\beta) P_{\gamma}(\tilde\gamma) \nonumber \\
    &P_A(a|\tilde\beta,\tilde\gamma) P_B(b|\tilde\gamma, \tilde\alpha) P_C(c|\tilde\alpha, \tilde\beta),
\end{align}
where $P_A(a|\tilde\beta \tilde\gamma)$ denoted the response function of Alice given some values of the local hidden variables $\tilde\beta$ and $\tilde\gamma$ that she receives from sources $\beta$ and $\gamma$, respectively, and similarly for Bob and Charlie. Note that we are currently interested in the discrete outcome case, in particular where $a,b,c \in \{1,2,3,4\}$.

In contrast to the classical correlations, having access to bipartite quantum sources allows one to sample from distributions of the form
\begin{align}\label{eq:quantum_model}
    P(a,b,c) = \text{Tr}\big(
    &\rho_{C_2,A_1} \otimes \rho_{A_2,B_1} \otimes\rho_{B_2,C_1}\cdot\notag\\
    \cdot &M_{A_1,A_2}^a \otimes M_{B_1,B_2}^b \otimes M_{C_1,C_2}^c
    \big),
\end{align}
where the subscripts denote the sub-Hilbert spaces each party, e.g. $\rho_{B_2,C_1}$ is the density matrix of the quantum state distributed by source $\alpha$, and $\{M_{A_1,A_2}^a\}_a$ is the Positive Operator-Valued Measure which describes Alice's measurement, such that $M_{A_1,A_2}^a \geq 0$ and $\sum_a M_{A_1,A_2}^a = \mathbb{I}$.

When a distribution $P$ has no classical explanation according Eq.~\ref{eq:local_model}, it is called triangle-nonlocal, or simply just nonlocal in the current context. A prominent example is the Elegant distribution~\cite{Gisin2019}, whose nonlocality was recently proven in Ref.~\cite{gitton_paper}, with supporting numeric and analytic studies for its noise robustness in Refs.~\cite{NNpaper, 2024_ineq_Tamas, silva2023numerical}. To sample from the distribution the source should distribute singlets, and the parties should measure in the basis
\begin{equation}\label{eq:EJM_Measurement}
    \ket{\Phi_i} = \frac{\sqrt{3}+1}{2\sqrt{2}}\ket{\vec{m}_i,-\vec{m}_i} + \frac{\sqrt{3}-1}{2\sqrt{2}}\ket{-\vec{m}_i,\vec{m}_i},
\end{equation}
where $i \in \{1,2,3,4\}$, $\vec{m_i}$ are vertices of a tetrahedron in the Bloch sphere, $\ket{\vec{m}_i}$ are the corresponding single qubit states, and $\langle\vec{m}_i|-\vec{m}_i\rangle = 0$. The distribution obtained from these so-called EJMs is symmetric both under permutation of the parties and of the outcomes, and can thus be characterized by 3 parameters (or 2 when considering normaliztion),
\begin{equation}\label{eq:EJM_Distribution}
    P_E (a,b,c) = \begin{cases}
\frac{25}{256} & a=b=c,\\
\frac{1}{256} & a=b \neq c \text{ or } b=c \neq a \text{ or } c=a\neq b,\\
\frac{5}{256} & a\neq b \neq c\neq a.
\end{cases}
\end{equation}

In general, deciding whether a local model exists is a difficult task. The set of local correlations forms a semi-algebraic set, and is characterized by polynomial inequalities. Exact characterization of this set can be constructed, however its computational cost is prohibitive~\cite{Lee2017}. Analytic relaxations of the local set that lead to Bell-type inequalities have been developed~\cite{Chaves2016,Chaves2021}, however these are too weak to prove the nonlocality of distributions such as the Elegant one. A systematic relaxation of the local set can be made via the inflation technique, shown to converge to the local set~\cite{InflationTechnic,Navascues2020,InflationImplementation}.
Despite the convergence, generating tight relaxations is numerically demanding, the limits of which were pushed in the recent proof by heavily relying on the symmetries of the Elegant distribution~\cite{gitton_paper}. Unfortunately, the certificate and the same numerics can not be practically applied to distributions which are not exactly, but only approximately symmetric, e.g. those that appear in real-world experiments such as ours. Deciding nonlocality by showing that no local model exists through an exhaustive inner search of the local set, is extremely expensive and is only feasibile for subsets of the local set of low-cardinality scenarios~\cite{daSilva2023,daSilva2025}. In order to perform more complete searches of the local set, neural network-based ans\"atze have been used (LHV-Net)~\cite{NNpaper}.  These are variational heuristics for finding local models, and do not give hard proofs of nonlocality. Nonetheless, in properly characterized scenarios they give reliable indications of (non)locality~\cite{pozas2023proofs,singlephotontriangle,Krivachy_closingdetectionloophole, towardsaminimalexample, toprobNNL, thesis_girardin}. This is the case for quaternary-outcome distributions such as the Elegant one, the neighborhood of which has been studied extensively both analytically and numerically~\cite{NNpaper,daSilva2023,2024_ineq_Tamas}. In the current work, we thus rely on the neural network technique and on the noise-robust, though not yet proven, Bell inequalities laid out in Ref.~\cite{2024_ineq_Tamas} to evaluate whether the experimentally observed distribution is nonlocal or not. Since it is central in the evaluation of our results, in the following we summarize how the neural network technique can be used to deduce that a distribution is nonlocal, and give further details in the Supplemental Material~\cite{SI}.

A feed-forward artificial neural network is a numeric model for any multivariate, multidimensional function. Its parameters can be fit (it can be trained), in order to minimize a differentiable objective function. The core idea of the neural network technique is to model each of the response functions in (\ref{eq:local_model}) with neural networks. For example, Alice's neural network would take as inputs some $\beta_i$, $\gamma_i$, and output a normalized vector $P_{A}^{\text{NN}}(a|\beta_i,\gamma_i)\in\mathbb{R}^4$.
Sampling over many ($M$) triples $(\alpha_i,\beta_i,\gamma_i)\in[0,1]^{3}$, one arrives at a Monte Carlo estimate of (\ref{eq:local_model}),
\begin{equation}
    P_{\text{NN}} = \frac{1}{M}\sum_{i=1}^{M} P_{A}^{\text{NN}}(a|\beta_i,\gamma_i) P_{B}^{\text{NN}}(b|\gamma_i, \alpha_i) P_{C}^{\text{NN}}(c|\alpha_i,\beta_i).
\end{equation}
Crucially, each party's neural network only has access to the respective hidden variables allowed by the triangle structure, thus any distribution given by the neural network is \textit{by construction} local. Due to the universal approximation properties of deep neural networks, by increasing the size of the network one can in principle parametrize the whole local set. The distribution $P_\text{NN}$ depends on the parameters of the neural networks, which are initialized randomly. One then trains the neural network via gradient descent techniques, minimizing the loss.

LHV-Net is often used to find the distance between a target distribution and the local set~\cite{NNpaper}, in which case we take the loss to be the Euclidean distance between the target distribution $P_\text{target}$ and $P_\text{NN}$, $||P_{\text{target}}-P_{\text{NN}}||_2$. Due to the constructive nature of the approach, any observed loss is an upper bound on the distance between the target distribution and the local set, in particular resulting in strong claims when the distance is small. In practice, the tool is often used by taking a target distribution and adding physical noise to it controlled by a parameter $v\in[0,1]$, which results in the target distribution becoming more and more local ($P_{\text{target}}^{v=0}$ being the most noisy, local distribution by convention). We then plot the smallest distance found by the neural network as a function of $v$ by retraining the neural networks from scratch, independently, for several $v$ values. In all cases we expect to see close to zero distance at $v=0$. If $P_{\text{target}} \equiv P_{\text{target}}^{v=1}$ is local, then the distance should stay at small values in the whole range of $v$. However, if it is nonlocal, then we should observe an increase in the distance at the $v$ value corresponding to the edge of the local set. We use this as one of the main criteria for deducing whether our observed experimental distribution is nonlocal or not.

Finally, notice that LHV-Net can be optimized to minimize (maximize) any differentiable loss function, and hence can also be used to test conjectures of Bell inequalities. The authors in Ref.~\cite{2024_ineq_Tamas} examined a candidate family of Bell inequalities tailored to the Elegant distribution. The inequality formalizes our intuition that local distributions in the triangle network can only become more correlated than the Elegant distribution if they are not symmetric. Specifically, the conjecture is that for local distributions $P$
\begin{equation}\label{eq:inequality}
    w \left[ s_{111}(P) - s_{111}(P_\text{E})\right] - (1-w) \Delta(P) + \delta_w \leq 0,
\end{equation}
where $s_{111}(P) = \sum_k P(k,k,k)$ captures the strength of the correlations, $\Delta(P)$ is a distance from the symmetric subspace (see Supplemental Material~\cite{SI} for the precise form), $w$ is a free parameter allowing a trade-off of how important of a role correlatedness and symmetry play, and $\delta_w$ is the smallest difference between the Elegant distribution's and any local model's $s_{111}$ values within the symmetric subspace. For any value of $w$, LHV-Net can be used to find the $\delta_w$ value. If this value is positive, then we have a conjecture for an inequality. For a wide range of $w$ values Ref.~\cite{2024_ineq_Tamas} finds a positive $\delta_w$, leading to inequalities which we will test with our experimental data. The great advantage of working with these inequalities is that they are interpretable and robust to noise and asymmetries. The shortcoming is, however, that despite these inequalities falling in line with our intuition and being reinforced by numerical searches, they are not proven (yet).

\begin{figure}
  \centering
  \includegraphics[scale = 0.35]{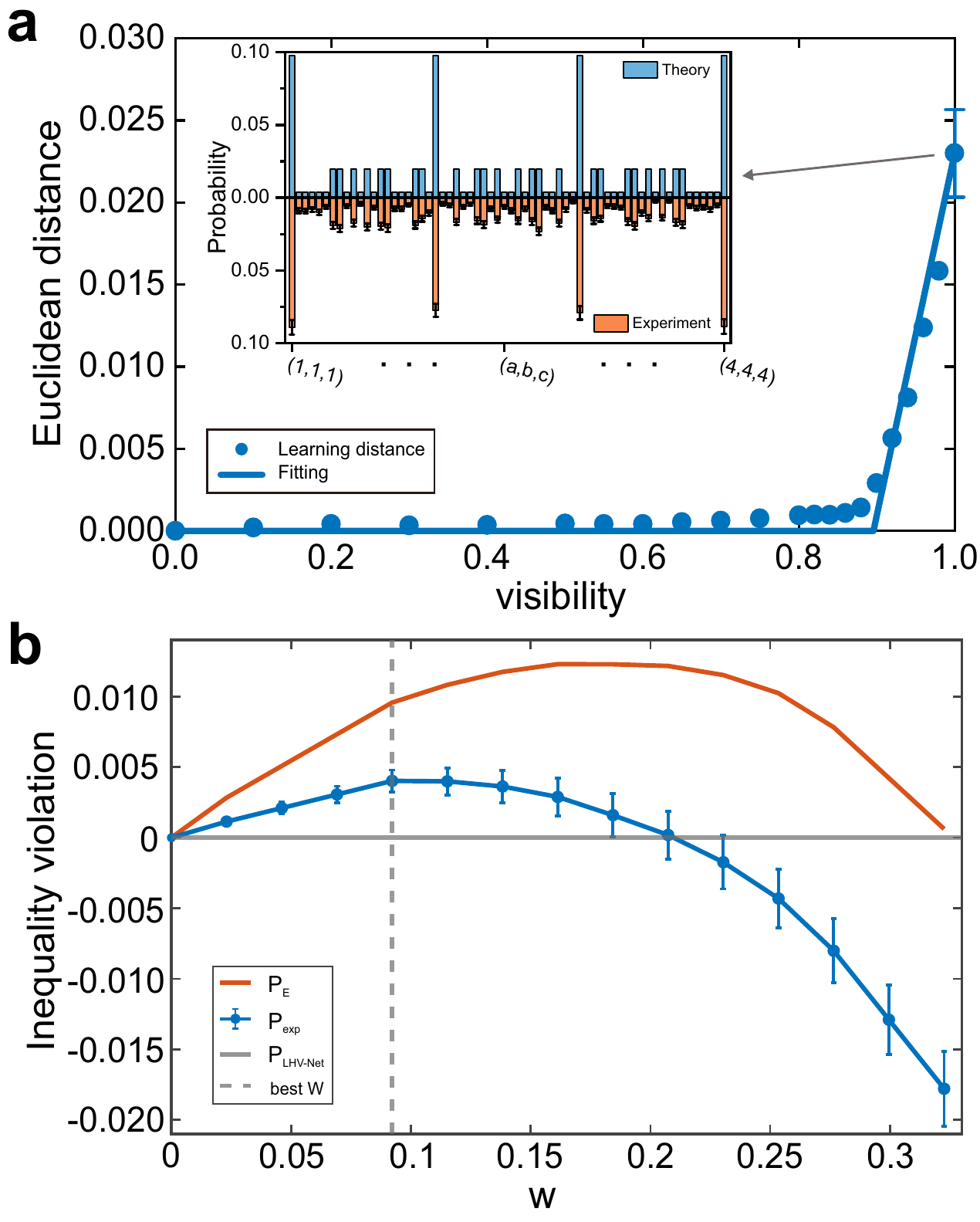}
  \caption{Experimental result. (a) Euclidean distance between the experimental distribution and the local set (determined by the neural network), plotted as a function of measurement visibility $\nu$. We add equal white noise to the measurement at each node, the measurement visibility is determined by the intensity of white noise, e.g., the measurement operator for Alice becomes $\nu M^a_{A_1,A_2}+\frac{1-\nu}{4}I$ (see Supplemental Material~\cite{SI} for details). The inset shows the theoretical and experimental Elegant distribution, the x-coordinate denotes the outcome for the three parties, the error bars are deduced from the photon statistical error. (b) Violations of Eq.~\ref{eq:inequality} for the theoretical (red) and experimental (blue) Elegant distribution as a function of $w$. The largest violation is observed at $w\approx0.09$ for experimental result. }
  \label{result}
\end{figure}

\emph{Optical triangle network.---} The triangle network is constructed from three optical Einstein-Podolsky-Rosen (EPR) sources, each located on one side of the triangle.
The EPR source generates entangled photon pairs through spontaneous parametric downconversion (SPDC) process, in a “sandwich-like” nonlinear crystal pumped by an ultrafast laser pulse. The produced polarization-entangled state is $\ket{\psi^-} = (\ket{HV}-\ket{VH})/\sqrt{2}$, with qubits encoded in the polarization degree of freedom of the photons. 
Each node of the triangle network receives two photons from two different EPR sources. The experimental sketch is shown in Fig.~\ref{setup}.

An important condition is that the EPR sources should be independent of each other. To improve the independence of the sources, we split the pulse from the single laser into three and let them pump the three EPR sources in parallel so that the photon pairs are generated from independent crystals. 
Then we insert a tiltable optical window in each pumping path. The tilt angle of the optical windows, which can change the optical distance, are controlled by independent quantum random number generators. 
In this way, the randomly tilted window imposes a completely random phase for each pump beam, thus erasing the coherent information between them. 
Such a method has been used in previous studies of network nonlocality~\cite{saunders2017experiment, NingNing2023experiment}.

The three nodes of the network all perform fixed EJM, thus there are no external inputs. We obtain the probability distribution of EJM by projecting the input states onto each of the four EJM bases separately.
We construct four projection settings in the experiment, each corresponding to one of the EJM bases. We switch these settings randomly during the experiment 
and finally combine the results to obtain the distribution under the fair sampling assumption.
As shown in Eq.~\ref{eq:EJM_Measurement}, the EJM bases are both partially entangled and have the same Schmidt coefficients. 
Our EJM basis projection device contains two parts. One is a partially entangled projection device that projects the input state into a partially entangled state with the same Schmidt coefficient as the EJM bases. 
This consists of a standard photonic polarization Bell state projection setup, and a customized polarization-dependent loss element that biases the Schmidt coefficients of the projected state to match the EJM bases. 
Another part is a basis transformer that enables the transformation between $\ket{H}/\ket{V}$ and the Schmidt bases $\ket{\vec{m}_i}/\ket{-\vec{m}_i}$ of EJM bases. 
This is achieved by three cascade wave plates mounted on motorized rotation stages. Switching between the four projection settings is accomplished only by rotating the cascade wave plates to the corresponding angles. (See more details in Supplemental Material~\cite{SI}).
Each node performs the EJM in this way, eventually normalizing the raw data we can obtain the Elegant distribution.

\emph{Experimental results.---} We first characterize the experimental setup. Our EPR sources achieve both high brightness (0.2 MHz) and high collection efficiency (31\%), and with quantum state tomography 
we find their fidelities (defined as $F = \bra{\psi^-}\rho\ket{\psi^-}$) to be $0.9769 \pm 0.0001$, $0.9761 \pm 0.0001$, and $0.9802 \pm 0.0001$, respectively. 
Similarly, we analyze the measurement setup at each node with measurement tomography, obtaining fidelities of $0.9352 \pm 0.0017$, $0.9384 \pm 0.0020$, and $0.9382 \pm 0.0019$, respectively.

During the experiment, we perform 64 measurement settings (each node has 4 settings) to obtain an Elegant distribution. To reduce the impact of laser power fluctuations on counting rates, 
we randomly switch the measurement settings every 10 minutes. Each setting is measured 27 times, and finally, we collect 3343 six-fold events in 288 hours. 
The optical windows randomly switch tilt angles every 20 milliseconds, which is much less than the time required to detect a six-fold event, thus the quantum coherence between the three pump beams 
is destroyed on the time scale of the network. We plot the experimental Elegant distribution $P_{\text{Exp}}$ against the theoretical distribution in the inset of Fig.~\ref{result}(a).

Since a technique to give a direct proof of nonlocality for our distribution does not exist, in order to determine whether our experimental Elegant distribution $P_{\text{Exp}}$ is compatible with a local model, we evaluate it with three separate techniques. Here we discuss the most direct one, the neural network search for local explanations, in detail. We leave the details of the other two, a hypothesis test and the evaluation of the inequality Eq.~\ref{eq:inequality}, to the Supplemental Material~\cite{SI} (Sections I.B and I.C, resp.).

For the distribution observed in the experiment as the target distribution, by independently running the neural network 60 times, 
we find that the minimal Euclidean distance between the local set and $P_{\text{Exp}}$ is $0.0230 \pm 0.0027$, 
where the uncertainty is determined by running 50 Monte Carlo simulations of the measured data based on Poisson distributed photon statistics and calculating the minimal distance of each set of the simulated data by running the neural network 20 times. The uncertainty represents one standard deviation and contains both the photonic statistical error and the error induced by the neural network.
We proceed to see whether a transition in the neural network's loss function occurs as we introduce noise artificially via measurement visibility, akin to the $v$ parameter described in the introduction. The results and a clear transition can be seen in Fig.~\ref{result}(b), indicating nonlocality of $P_{\text{Exp}}$. We fit the points with visibility from 0.9 to 1 into a line. The line intersects the x-coordinate at about 0.89 which we treat as the critical visibility. Due to the limited numerical precision of the neural network and its training, it is not possible for it to return a local model that exactly replicates the noise distribution even below the critical point, so the distance will not be exactly 0 but rather a very small value. Our experimental Elegant distribution $P_{\text{Exp}}$ has a distance to the local set that is more than 8 standard deviations larger than the distance at the critical point, further strengthening the case of nonlocality and of the robustness of the evaluation.

Finally, let us summarize the results of the other two techniques that we study in the Supplemental Material. We show in Section I.B that a significant portion, about 13\%, of the Euclidean distance map presented in Ref.~\cite{2024_ineq_Tamas} would be different \textit{if} $P_{\text{Exp}}$ would actually be local. Moreover, in Section I.C and also Fig.~\ref{result}(b) we show that the observed distribution $P_{\text{Exp}}$ can violate the inequality Eq.~\ref{eq:inequality} by more than 5 standard deviations (see Fig.~\ref{result}b). All three techniques confidently and robustly indicate that the implemented distribution $P_{\text{Exp}}$ has no local explanation. We envision that in future work our data can be evaluated using more advanced techniques in order to give a definitive proof of nonlocality. Our work motivates further work in developing such methods.

\emph{Discussion.---} The experimental implementation of the Elegant distribution motivates further research into this topic, shifting the focus towards applications. Nonlocality in quantum networks was already used for self-testing~\cite{Pavel2023, toprobNNL}, semi-device independent randomness generation~\cite{Sarkar} and distributed computing~\cite{Balliu2025}.
Yet, most of these applications are based on restricted types of quantum measurements (e.g., separable or Bell state measurements). We expect that beyond these, new types of quantum measurements such as the  elegant measurement, already proven to be optimal for certain tasks~\cite{CzartowskiZyczkowski2021}, will lead to new potential applications of network nonlocality, as discussed in~\cite{MunshiPan2025}, as well as their experimental demonstration. This could solved some issues faced by currently considered measurements, and maybe prove new quantum advantages for certification and distributed computing tasks, overcoming some existing limits~\cite{CoiteuxRoy2024,Akbari2025}. 

Note however that our experiment is subject to the common loopholes in the standard Bell experiments, namely the locality loophole and the fair sampling loophole. 
In addition, the network local model also opens a new source independence loophole. More precisely, in the triangle network, the three distributed quantum sources should be independent.
However, just like the freedom of choice loophole in the standard Bell test, no argument can prove that this independence fully holds.
This condition can only be made more stringent, but the associated independence loophole, explaining the obtained correlations through correlated sources (which can then be distributing a LHV model) can never be closed. 
In this work, we experimentally enhance the source independence by erasing the coherence information between the pump beams of the sources, similarly to what has been used in several previous studies.

Contrary to previous experiments of triangle network nonlocality implementing the Fritz distribution~\cite{Polino2022experiment}, the Elegant distribution generated in our experiment is thought to be genuine to the triangle network, in the sense that it cannot be interpreted as reproducing a standard forms of Bell nonlocality embedded in the triangle scenario. 
In particular it relies on the use of entangled measurements at each party, while the Fritz's model can be realized by only separable measurements. Although Bäumer et al. also studied the Elegant distribution using a superconducting quantum computer, they didn't provide evidence that the produced Elegant distribution is nonlocal~\cite{baumer2021}. We discuss further aspects of this in Supplemental Material III~\cite{SI}. In contrast, the high fidelities of the entangled sources (EJMs) of $97.8\%$ ($93.7\%$) of our experiment allow us to demonstrate nonlocality in the triangle network.

Another feature that differentiates the Elegant model to the standard Bell scenario is that the Elegant distribution can be generated by fixed measurements without external inputs. 
Although in the experiment we didn't achieve an deterministic EJM, we just project the input states onto each EJM basis separately, the distribution can be obtained in principle simultaneously. This unideal realization of EJMs may introduce the freedom of choice loophole like the standard Bell test. By using auxiliary entangled photons, it is able to implement a four-output EJM. Although other platforms (superconducting, ions) can perform deterministic entangled measurements with high fidelity, photonic system is more realistic for network nonlocality tests than other systems as future quantum networks will concern distant parties and photons to be the most appropriate information carriers for distant communications.

\emph{Acknowledgments.---} This work was supported by the Innovation Program for Quantum Science and Technology (No. 2021ZD0301604). The numerical calculation is partially implemented in the Supercomputing Center of University of Science and Technology of China. NG acknowledges the Swiss National Science Foundation via the NCCR-SwissMap. TK acknowledges funding by the Austrian Federal Ministry of Education, Science and Research via the Austrian Research Promotion Agency (FFG) (flagship project  FO999897481 funded by the European Union – NextGenerationEU), as well as funding by the Swiss National Science Foundation (project P500PT\_214458) and by CEX2019-000910-S [MCIN/ AEI/10.13039/501100011033], Fundació Cellex, Fundació Mir-Puig, and Generalitat de Catalunya through CERCA. MOR acknowledges funding by the ANR for the JCJC grant LINKS (ANR-23-CE47-0003) and T-ERC QNET (ANR-24-ERCS-0008), and by the European Union’s Horizon 2020 Research and Innovation Programme under QuantERA Grant Agreement no. 731473 and 101017733.

\clearpage

\begin{widetext}
    \section*{Supplementary materials}

\section{Techniques for demonstrating nonlocality}
\subsection{Neural network search for local hidden variable models}

We use the best available numeric methods to reinforce that our experimental results could not have been obtained from a classical model according to Eq.1 in the main text
\begin{align}\label{eq:local_model}
    P(a,b,c) = &\int d\alpha d\beta d\gamma P_{\alpha}(\alpha) P_{\beta}(\beta) P_{\gamma}(\gamma) \nonumber \\
    &P_A(a|\beta,\gamma) P_B(b|\gamma, \alpha) P_C(c|\alpha, \beta),
\end{align}
Specifically, we use the neural network-based ansatz developed in Ref.~\cite{NNpaper}, (LHV-Net). In this, the authors show that modeling local hidden variable models with artificial neural networks is a reliable heuristic, reproducing benchmark results, as well as providing new conjectures, which have since been partially proven~\cite{pozas2023proofs}.

Here we give more details on the design and implementation of the neural networks. The classical network LHV model can be represented by a directed acyclic graph (DAG) that illustrates its causal structure, while a feedforward neural network can also be depicted by a DAG. This allows a direct encoding of a network LHV model into neural networks. Fig.~\ref{fig:PNN}(a) shows the DAG of the triangle network, in this scenario, the three sources sample independent random variables $\alpha$, $\beta$, $\gamma$ to the three parties according to the triangle causal structure, and the three parties process the inputs through their local response functions and output a number $a,b,c\in\{1,2,3,4\}$ respective. Fig.~\ref{fig:PNN}(b) provides an equivalent structure of Fig.~\ref{fig:PNN}(a), where the three sources are positioned at the top, and information flows from top to bottom. Fig.~\ref{fig:PNN}(c) illustrates the structure of encoding the triangle network into a neural network. The local hidden variables of the three sources serve as the input layer. The hidden layer is divided into three blocks, each corresponding to the local response function of the party. According to the causal structure of the triangle network, each block is only connected to its corresponding sources. By encoding the neural network like this, the generated distribution is guaranteed to be from the local set (Eq.~\ref{eq:local_model}) by construction.

The loss function can be any differentiable measure of discrepancy between the target distribution $P_{\mathrm{target}}$ and the distribution generated by the neural network $P_{\mathrm{NN}}$. Here we employ the Euclidean distance as the loss function, defined as $d(P_{\mathrm{target}},P_{\mathrm{NN}}) = \sqrt{\Sigma_i{|P_{\mathrm{target}}(i)-P_{\mathrm{NN}}(i)|^2}}$.
Since the neural network will never exactly reproduce the target distribution due to finite number of samplings, our method to deciding nonlocal distributions is to search for qualitative changes in the machine’s behavior when transitioning from the nonlocal to the local set.

We employed a fully connected multilayer neural network with a depth of 4 layers and a width of 20 for each block. The input consists of uniformly distributed hidden variables with a batch size of 8000. The weights are iteratively updated 20000 times to optimize the loss function, and a total of 48 epochs are completed for one round of neural network training. In order to speed up training, in the first few epochs we use the Kullback-Leibler divergence as the loss, defined as $D(P_\text{target}||P_\text{NN})=\sum_i P_\text{target}(i)\log\frac{P_\text{target}(i)}{P_\text{NN}(i)}$. The weight update is done with Adadelta at first, followed by stochastic gradient descent with a decreasing learning rate for fine-tuning. The code for the neural network is available on the GitHub~\cite{code}. We run the program on the serveres of the Supercomputing Center of University of Science and Technology of China.

For the target distribution obtained experimentally, we run the program 60 times to find the minimum distance to the local set. If the distribution could be explained by a classical (local hidden variable) model, we expect the neural network would find it and the resulting objective would reach zero. As portrayed in the Results section of the main text, the closest local model the neural network could find is $0.0230 \pm 0.0027$, well above the zero value. In contrast, in Ref.~\cite{NNpaper}, the authors find a distance of approximately $0.05$ for the $P_E$, the theoretical Elegant distribution. Our experimental results are at about $50\%$ distance between the noiseless Elegant distribution and the local set. To calculate the errorbar, we use Monte Carlo simulation to generate 50 sets of data based on the raw data and Poisson distributed photon statistics. For each set of data, we run the program 20 times to calculate the minimum distance. We calculate the standard deviation of these 50 results to obtain the errorbar. Then we add artificial white noise to the measurement process at each node, controlled by the parameter of measurement visibility $\nu$. The noise distribution can be calculated according to Eq.~\ref{white_noise}. We choose $\nu=\{0, 0.1, 0.2, 0.3, 0.4, 0.5, 0.55, 0.6, 0.65, 0.7, 0.75, 0.8, 0.82, 0.84, 0.86, 0.88, 0.9, 0.92, 0.94, 0.96, 0.98, 1\}$. For each noise distribution we run the program 20 times to find the minimum distance. We also utilize the Euclidean distance to analyze the noise robustness, as it typically results in a straight line in the nonlocal region, which helps guessing the visibility where the distribution becomes local (see Fig.3(a) in the main text).

\begin{figure}
  \centering
  \includegraphics[width=\textwidth]{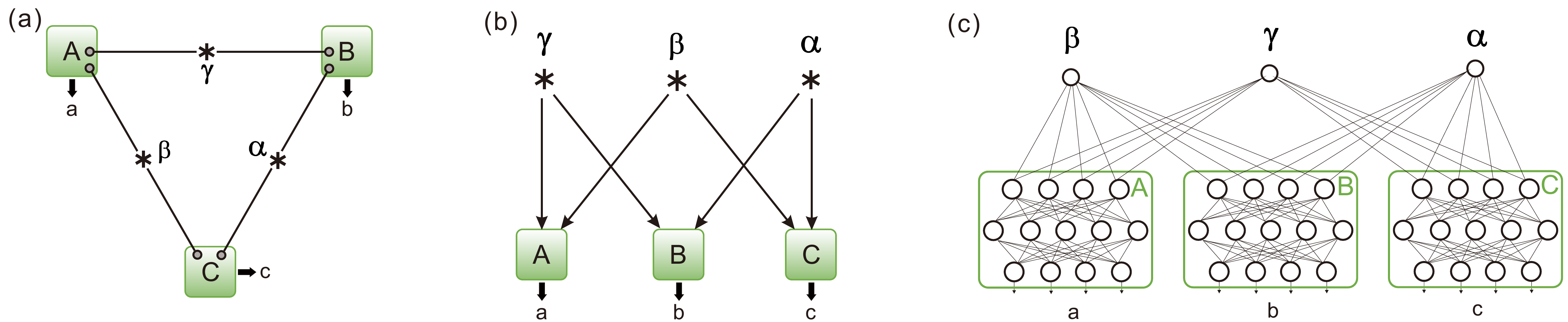}
  \caption{(a) The DAG of the classical triangle network. (b) An equivalent description of the triangle network. (c) The configuration of the neural network for the triangle network.  }
  \label{fig:PNN}
\end{figure}

\subsection{Symmetric subspace map distortion}
Furthermore, in Ref.~\cite{2024_ineq_Tamas}, the authors run the neural network search independently for many points in the symmetric subspace, creating a map of distance to the local set (as probed by the neural network) on which local regions are clearly drawn out in dark blue (reproduced in Fig.~\ref{fig:LHV_net_map_comparison} (a))  - this map being in excellent correspondence with analytic expectations (see Ref.~\cite{2024_ineq_Tamas} for details). We asked ourselves: if the observed distribution $P_{\text{Exp}}$ would be local, how would this map look like? Ineed, as displayed in Fig.~\ref{fig:LHV_net_map_comparison}, in a significant portion of the symmetric subspace (in about 34\% of the semmingly nonlocal part above the top part), the neural network should have found something closer than what it actually did \text{if} $P_{\text{Exp}}$ would be local. We believe that such a strong deviation between what we observed in the neural network's scans (Fig.~\ref{fig:LHV_net_map_comparison} (a)) and what would be the case if $P_{\text{Exp}}$ would be local (Fig.~\ref{fig:LHV_net_map_comparison} (b)) further strengthen the case that $P_{\text{Exp}}$ is nonlocal. Indeed, if this were not the case it would mean that in this red region the neural network scan missed $P_{\text{Exp}}$ and nearby distributions in each case.

\begin{figure}
  \centering
  \includegraphics[width=\textwidth]{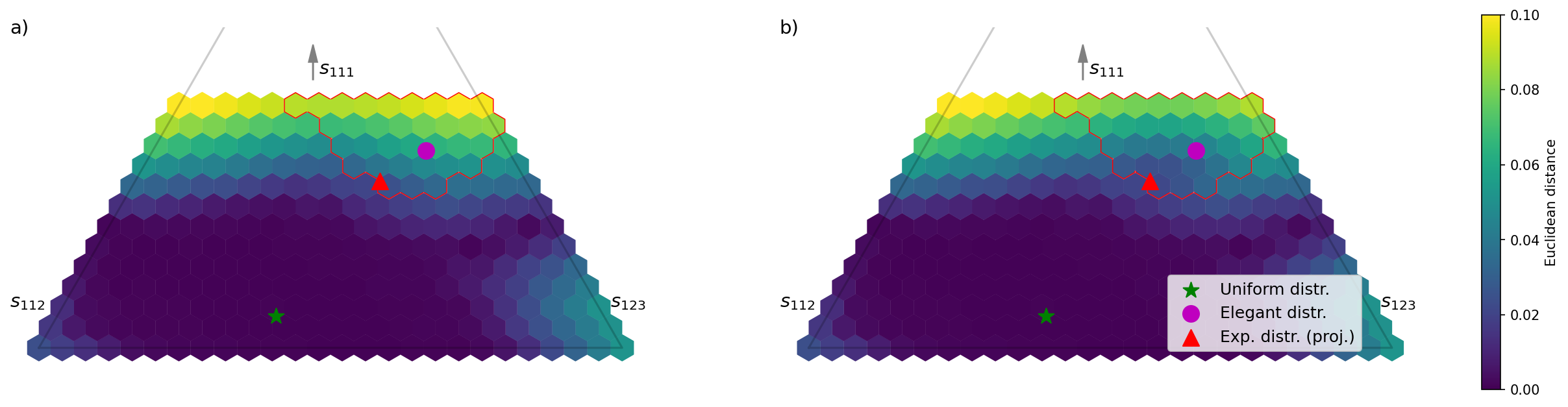}
  \caption{a) Smallest Euclidean distance that the neural network could find to 260 points in the symmetric subspace (cut off at 0.1, reproduced from Ref.~\cite{2024_ineq_Tamas}). Significant points such as the uniform and Elegant distribution are depicted, as well as the projection of $P_{\text{Exp}}$ to the symmetric subspace. b) What the neural network should have found if the experimentally observed $P_{\text{Exp}}$ would be local. Generated by taking the original neural network scan; then we calculated the distance $P_{\text{Exp}}$ to each of the 260 points and if this distance is smaller than what the neural network found, it replaces the neural network's distance. Target distributions where $P_{\text{Exp}}$ is closer than what the neural network found are within the red boundary.}
  \label{fig:LHV_net_map_comparison}
\end{figure}

\subsection{Violation of conjectured Bell inequality}
Finally, note that LHV-Net, together with analytic considerations, has been used to derive inequalities, by changing the objective function from a distance function to an inequality~\cite{2024_ineq_Tamas}. The inequality captures the trade-off between the strong correlations and the strictness of the symmetry constraint. In detail, one can define the function
\begin{align}
    f_w(p) = w \cdot s_{111}(P) - (1-w) \Delta(P),
\end{align}
where
\begin{equation}\label{eq:s111def}
    s_{111}(P) = P(a=b=c) = \sum_k P(k,k,k)
\end{equation}
captures the strength of the correlations, while $\Delta$ gives a penalty for being non-symmetric by summing up the deviations from the mean of each of the 3 types of events ((1,1,1); (1,1,2); and (1,2,3) -type outcomes), via
\begin{align}\label{eq:deltadef}
\Delta(P) =& \sum_{X \in \{111,112,123\}} \sum_{\{a,b,c\} \in \mathcal{I}_X} |M_X - P(a,b,c)|^2,\\
M_{X} =& \frac{1}{|\mathcal{I}_X|} \sum_{\{a,b,c\} \in \mathcal{I}_X} P(a,b,c),
\end{align}
where $\mathcal{I}_X$ is the index set of $X$-type outcomes (in particular $\mathcal{I}_{111}$ will contain 4 elements, $\mathcal{I}_{112}$ 36, and $\mathcal{I}_{123}$ 24 elements). Note that this limit in the strength of (1,1,1)-type correlations is visibile in the map in Fig.~\ref{fig:LHV_net_map_comparison}, where dark blue (local) points only appear on the bottom part of the symmetric subspace.

Then one compares the maximum that this function takes over LHV models, and compares it to the value that $P_E$ achieves,
\begin{align}
    \delta_w := f_w(P_E) -  \max_{P\in\mathcal{L}}f_w(P),
\end{align}
where $\mathcal{L}$ is the set of distributions admitting an LHV model. If $\delta_w>0$, then the inequality for that $w$ is a Bell inequality which certifies the nonlocality of $P_E$,
\begin{align}
    f_w(P) \leq f_w(P_E) - \delta_w.
\end{align}
Using the numeric tools of LHV-Net, one can obtain an estimate of the $\delta_w$ values, as was done in Ref.~\cite{2024_ineq_Tamas}.
In Fig.~\ref{fig:inequality} we plot $f_w(P) - f_w(P_E) + \delta_w$, which should be negative or 0 for all local distributions.  From the results we deduce that the inequality which most strongly certifies our experimental distribution is for $w\approx0.0922, \,\delta_w \approx 0.00957$, resulting in the inequality of
\begin{equation}\label{eq:ineq_l2}
    0.0922 \, s_{111}(P) - 0.9078 \,\Delta(P) \leq 0.0264,
\end{equation}
We identified this $w$ value by finding the largest ratio between the violation of $P_{\text{Exp.}}$ and $P_E$, which was 42\%. For the theoretical distribution $P_E$, the optimal $w$ value is $w\approx 0.16$. For our experimental distribution, $s_{111}(P_{\text{Exp.}}) \approx 0.334,\, \Delta(P_{\text{Exp.}}) \approx 0.000409$, whereas for the theoretical distribution $s_{111}(P_{\text{E}}) = \frac{100}{256} \approx 0.3906,\, \Delta(P_{\text{E}}) = 0$.

\begin{figure}
  \centering
  \includegraphics[scale = 0.9]{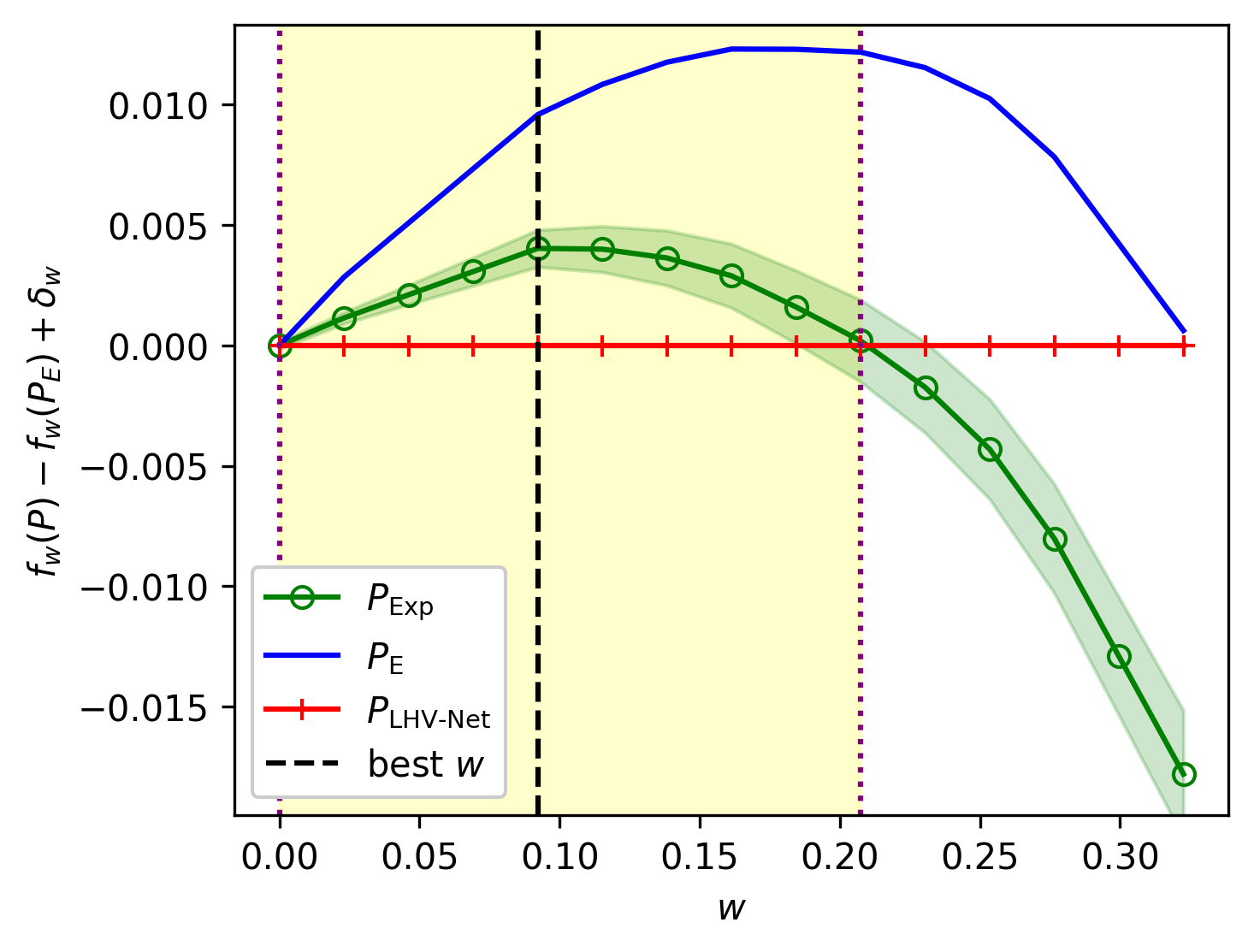}
  \caption{Violations of the conjectured inequality as a function of $w$, for the ideal EJM distribution (blue), the experimental results (green), and for classical strategies found by LHV-Net (red). Green band denotes one standard deviation, and yellow area displays $w$ values for which the inequality is violated by the experimental data. The strongest violation is observed at $w \approx 0.09$, reaching about 42\% of the ideal distribution's violation.}
  \label{fig:inequality}
\end{figure}

\begin{figure}
  \centering
  \includegraphics[scale = 0.7]{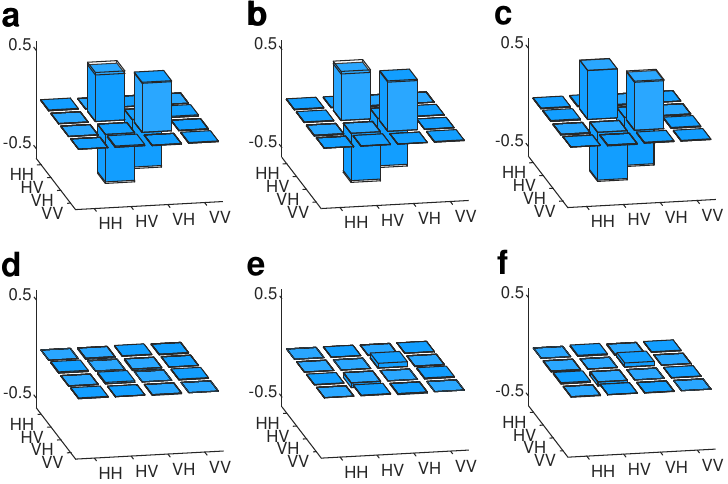}
  \caption{Tomographic results of the three EPR sources. a, b, and c are the real parts of $S_\alpha$, $S_\beta$, and $S_\gamma$ respectively. d, e, and f are the imaginary parts of $S_\alpha$, $S_\beta$, and $S_\gamma$ respectively. }
  \label{source_fidelity}
\end{figure}

\section{Experimental details}

\begin{figure}[H]
    \centering
    \includegraphics[scale=1.2]{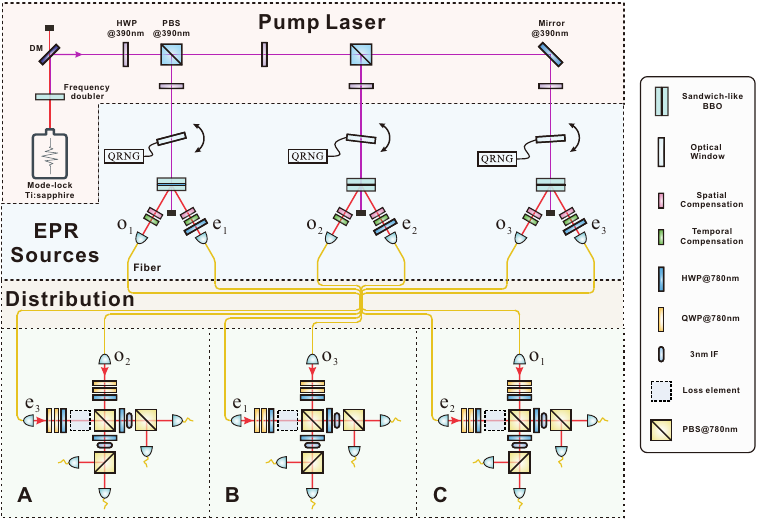}
    \caption{The detailed experimental setup. }
    \label{detail_setup}
\end{figure}

{\it Experimental setup.---}
Fig.~\ref{detail_setup} shows the detailed experimental setup, which consists of four parts from top to bottom: the preparation and distribution of the ultraviolet pulses, the EPR sources, the fiber distribution, and the measurement processes. The ultrafast laser pulses generated by the mode-locked Ti:sapphire laser (with a central wavelength of 780 nm, a pulse duration of 140 fs, and a repetition rate of 80 MHz) are first passed through a frequency doubler. The output ultraviolet laser is split into three beams averagely by using a series of HWPs and PBSs,
The relative phases between the beams are erased by inserting randomly rotated optical windows before they hit on the sandwich-like BBO crystals. Then the produced downcoversion photons are distributed to the three nodes by optical fibers, each of which receives one ordinary photon and one extraordinary photon from two different sources. In the measurement device, both photons pass through three cascaded wave plates, the extraodinary photon also passes through a polarization-dependent loss element. Then the two photons are overlapped on the central PBS for HOM interference. Then the two photons are measured in $|\pm\rangle$ basis, where $|\pm\rangle=(|0\rangle\pm|1\rangle)/\sqrt{2}$. When both output ports detect one photon and have the same polarization, the projection device succeeds. When all three projection devices succeed, we record a six-fold coincidence event.

{\it EPR source.---} The production of entangled photons is based on the spontaneous parametric downconversion (SPDC) process pumped by ultraviolet pulses. 
The pump pulses with a central wavelength of 390 nm are obtained from the frequency doubling system, where the fundamental frequency laser pulses 
are generated by the mode-locked Ti:sapphire laser with a center wavelength of 780 nm, a duration of 140 fs, and a repetition rate of 80 MHz. 
The EPR source is a sandwich-like structure composed of a true-zero-order half-wave plate (THWP) between two beta barium borate (BBO) crystals. 
Both BBO crystals are 2 mm thick and are identically cut for beam-like Type-II phase matching. When the ultraviolet laser pulse is incident on the crystal, 
both BBO crystals probabilistically produce a pair of extraordinary (e) and ordinary (o) photons with horizontal and vertical polarizations, i.e., $\ket{HV}$. 
The photons generated by the first BBO, however, will have their polarization flipped by the middle THWP, resulting in a state $\ket{VH}$. 
Then after temporal and spatial compensation, the two SPDC processes become indistinguishable, and the two-photon state becomes a singlet $\ket{\psi^-} = (\ket{HV} - \ket{VH})/\sqrt{2}$. 

In the experiment, we construct three EPR sources, each pumped with 340 mW ultraviolet pulses. When using a 3 nm spectrum filter for each side of the collection, 
each source has a counting rate of approximately 0.2 MHz, and the collection efficiency is approximately $31\%$. To characterize these EPR sources, we perform state tomography 
for each of them. The reconstructed density matrices are shown in Fig.~\ref{source_fidelity}, and the fidelity of the state $F = \bra{\psi^-}\rho\ket{\psi^-}$ is calculated to be 
$0.9769 \pm 0.0001$, $0.9761 \pm 0.0001$, and $0.9802 \pm 0.0001$, respectively.

{\it Source independence.---} The three sources are each pumped by three parallel pulses split from a single pulse. In each of the pumping paths, we insert a tiltable optical window, 
a 5 mm thick N-BK7 glass slice mounted on a motorized rotation stage, to impose an additional phase to the pulse. When the optical window deviates from the normal incidence angle by $0.6^o$, 
the pulse will have an additional optical distance of approximately 410 nm. During the experiment, three different quantum random number generators (QRNGs) control the tilt angle of the three optical windows. 
Each QRNG generates random numbers between 0 and 1, which are then mapped to angles with a resolution of $0.01^o$ between $0^o$ and $0.6^o$ to 
tilt the optical window, thereby applying random phase shifts between 0 and $2\pi$ to the pump beam. This process is repeated approximately every 20 ms, 
which is much faster than our six-fold counting rate of approximately $12$ per hour. Thus, the relative phase information of the three pump beams is effectively erased. 

\begin{figure}
  \centering
  \includegraphics[scale = 0.7]{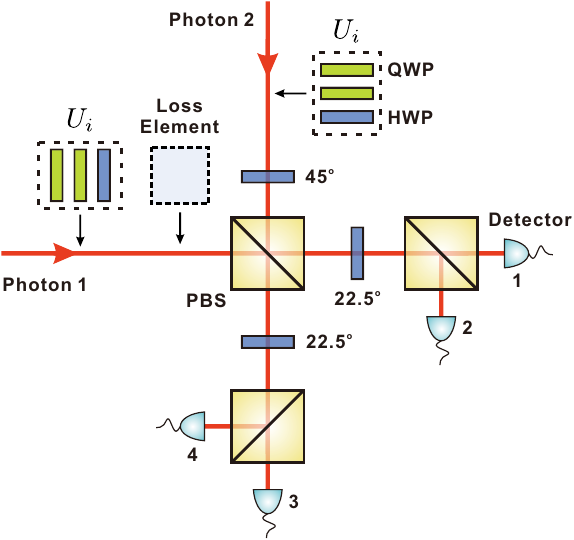}
  \caption{EJM basis projection device. The transmittance of the loss element to horizontal polarized photon is $T_H = 1$, while to vertical polarized photon is $T_V = 7-4\sqrt{3}$. 
  Each detector is preceded by an interference filter not shown. }
  \label{EJM_setup}
\end{figure}

{\it Elegant joint measurement.---} The EJM basis projection device is shown in Fig.~\ref{EJM_setup}. There is a standard Bell state projector consisting of a polarization beam splitter (PBS) 
with a $45^o$ half-wave plate (HWP) on the photon 2 path and two $22.5^o$ HWPs at each output of the PBS, the input state will be projected onto $(\ket{HV}+\ket{VH})/\sqrt{2}$ 
when the two output photons are located in the paths of detectors 1 and 3 or 2 and 4 respectively. The transformation made by the $45^o$ HWP can be absorbed into the unitary transformation. 
To match the Schmidt coefficients of the EJM bases, which are biased, we insert a polarization-dependent loss element in the path of photon 1 that is fully transmissive to $\ket{H}$ photon 
but has a transmittance of $7-4\sqrt{3}$ to $\ket{V}$ photon. 
By attenuating some of the vertical-polarized photons, the input state is projected toward a partially entangled state $\sqrt{\eta}(\frac{\sqrt{3}+1}{2\sqrt{2}}\ket{HV} + \frac{\sqrt{3}-1}{2\sqrt{2}}\ket{VH})$. 
Due to the introduction of the loss element, the projection efficiency becomes $\eta = 2(2-\sqrt{3})$. 
The two quarter-wave plates (QWPs) and one HWP at each input act as a basis transformer, which performs the unitary $U_i$ for each photon where $U_i\ket{\vec{m_i}} = \ket{H}$ and $U_i\ket{-\vec{m_i}} = \ket{V}$, 
so that the $\ket{H}/\ket{V}$ bases of the Bell state projector and the Schmidt bases $\ket{\vec{m}_i}/\ket{-\vec{m}_i}$ of the EJM basis can be converted. 
These wave plates are mounted on motorized rotation stages and can be rotated to specific angles to realize the conversion of any four sets of Schmidt bases of the EJM. 
Overall, the setup can project the input state toward any of the EJM bases $\sqrt{\eta}(\frac{\sqrt{3}+1}{2\sqrt{2}}\ket{\vec{m}_i, -\vec{m}_i} + \frac{\sqrt{3}-1}{2\sqrt{2}}\ket{-\vec{m}_i, \vec{m}_i})$, $i = 1, 2, 3, 4$. 
Note that the projection efficiency is the same for each projection setting, and the experimental results do not need to be renormalized for efficiency. 

\begin{figure}
  \centering
  \includegraphics[scale = 0.45]{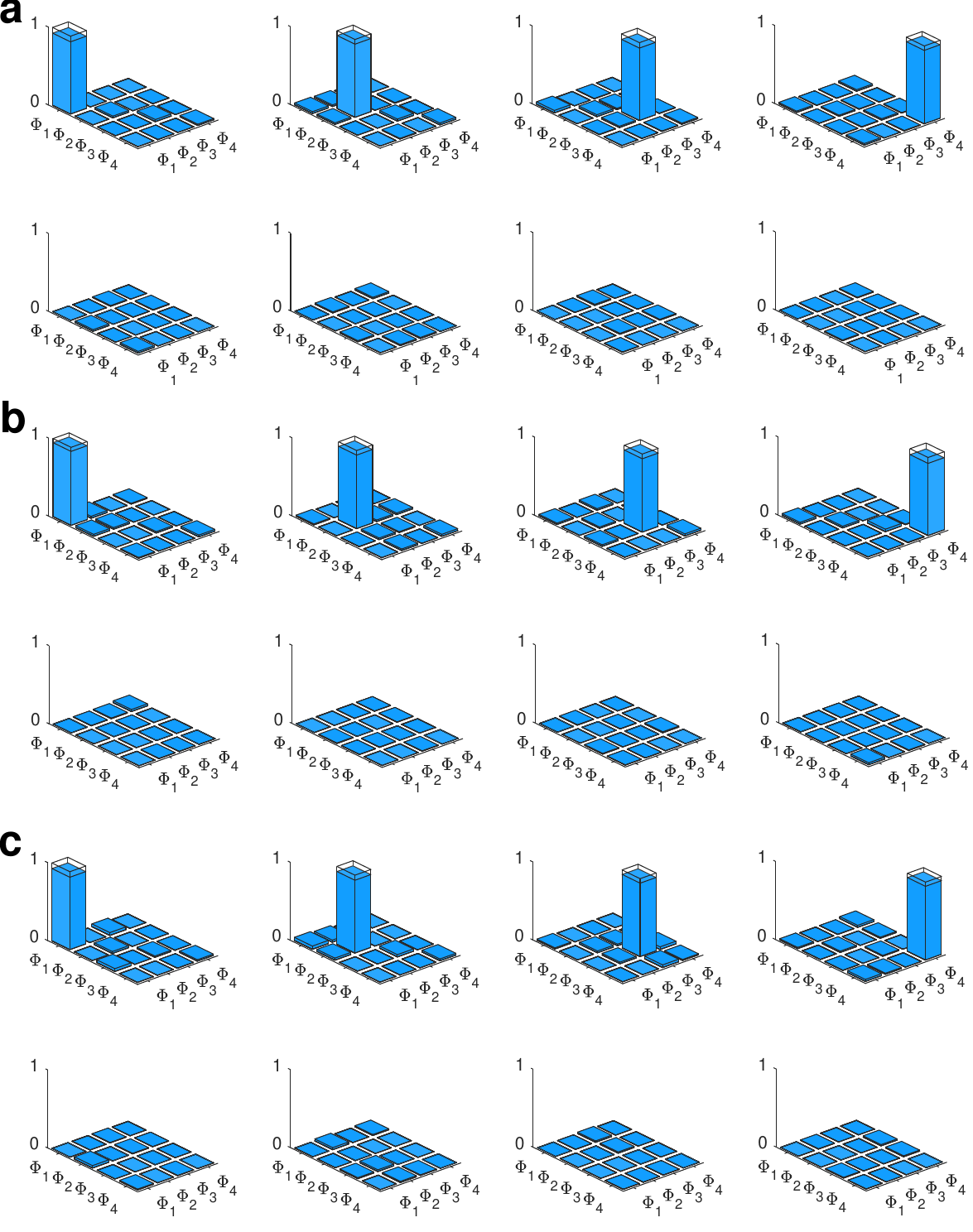}
  \caption{Tomographic results of EJM on three nodes. a, b, and c contain the reconstructed POVM elements on nodes A,B, and C, respectively, which are shown in EJM bases $\{\Phi_1, \Phi_2, \Phi_3, \Phi_4\}$. 
  In each subfigure, the top four are the real parts of the reconstruction matrices and the bottom four are the corresponding imaginary parts. }
  \label{EJM_fidelity}
\end{figure}

By means of the measurement tomography, we obtain the reconstructed matrices of the measurements at 
the three nodes with fidelity of $0.9352 \pm 0.0017$, $0.9384 \pm 0.0020$, and $0.9382 \pm 0.0019$. 
The results are shown in Fig.~\ref{EJM_fidelity}. 
Since the measurement setup is mainly affected by white noise, the original projectors became a series of Positive Operator-Valued Measure (POVM) elements in the experiment.

{\it Measurement visibility.---} The measurement visibility is determined by the intensity of white noise. Considering that our experimental measurement process suffers from more white noise, 
that is the POVM elements at nodes $A$, $B$ and $C$ become $\nu E^A_i+\frac{1-\nu}{4}I^A$, $\nu E^B_j+\frac{1-\nu}{4}I^B$, and $\nu E^C_k+\frac{1-\nu}{4}I^C$, where $i, j, k \in \{0, 1, 2, 3\}$ 
depend on the measurement output, and $\sum_{i=0}^{3}E^A_i = I^A$, $\sum_{j=0}^{3}E^B_j = I^B$, $\sum_{k=0}^{3}E^C_k = I^C$. The measurement visibility $\nu$ is between 0 and 1, 
$\nu = 1$ denotes our experimental measurement process, and $\nu = 0$ indicates that the measurement output is completely random. Based on this, we can calculate the 
noise distribution $P_\nu(i,j,k)$ at arbitrary measurement visibility using the following equation:

  \begin{align}
    \centering
    \label{white_noise}
    P_\nu(i,j,k) &= \mathrm{Tr}\big[\rho(\nu E^A_i+\frac{1-\nu}{4}I^A)(\nu E^B_j+\frac{1-\nu}{4}I^B)(\nu E^C_k+\frac{1-\nu}{4}I^C)\big] \nonumber \\ 
              &= \nu^3\mathrm{Tr}\big[\rho E^A_iE^B_jE^C_k\big] + \nu^2\frac{1-\nu}{4}(\mathrm{Tr}\big[\rho E^A_iE^B_jI^C)\big] + \mathrm{Tr}\big[\rho E^A_iI^BE^C_k)\big] + \mathrm{Tr}\big[\rho I^AE^B_jE^C_k)\big]) \nonumber \\ 
              &\quad \ + \nu(\frac{1-\nu}{4})^2(\mathrm{Tr}\big[\rho E^A_iI^BI^C)\big] + \mathrm{Tr}\big[\rho I^AE^B_jI^C)\big] + \mathrm{Tr}\big[\rho I^AI^BE^C_k)\big]) + (\frac{1-\nu}{4})^3\mathrm{Tr}\big[\rho I^AI^BI^C)\big] \nonumber \\ 
              &= \nu^3P_{\text{Exp}}(i,j,k) + \nu^2\frac{1-\nu}{4}(\sum_{k^\prime=0}^{3}P_{\text{Exp}}(i,j,k^\prime) + \sum_{j^\prime=0}^{3}P_{\text{Exp}}(i,j^\prime,k) + \sum_{i^\prime=0}^{3}P_{\text{Exp}}(i^\prime,j,k)) \nonumber \\
              &\quad \ + \nu(\frac{1-\nu}{4})^2(\sum_{j^\prime,k^\prime=0}^{3}P_{\text{Exp}}(i,j^\prime,k^\prime) + \sum_{i^\prime,k^\prime=0}^{3}P_{\text{Exp}}(i^\prime,j,k^\prime) + \sum_{i^\prime,j^\prime=0}^{3}P_{\text{Exp}}(i^\prime,j^\prime,k)) + (\frac{1-\nu}{4})^3
  \end{align}

where $P_{\text{Exp}}$ is the experimental elegant distribution. Feeding it into a neural network, we can learn how the measurement visibility $\nu$ affects the distance from the distribution to the local set.

{\it Raw data.---}
Fig.~\ref{rawdata} shows the raw data of the six-fold coincidence events for the 64 measurement settings. Each output c of node C corresponds to a panel. The vertical axis of each panel corresponds to the output a of node A and the horizontal axis corresponds to the output b of node B. By normalizing the 64 raw data, we can get the experimental elegant distribution $P_{\text{Exp}}(a,b,c)$.

\begin{figure}[H]
    \centering
    \includegraphics[scale=0.8]{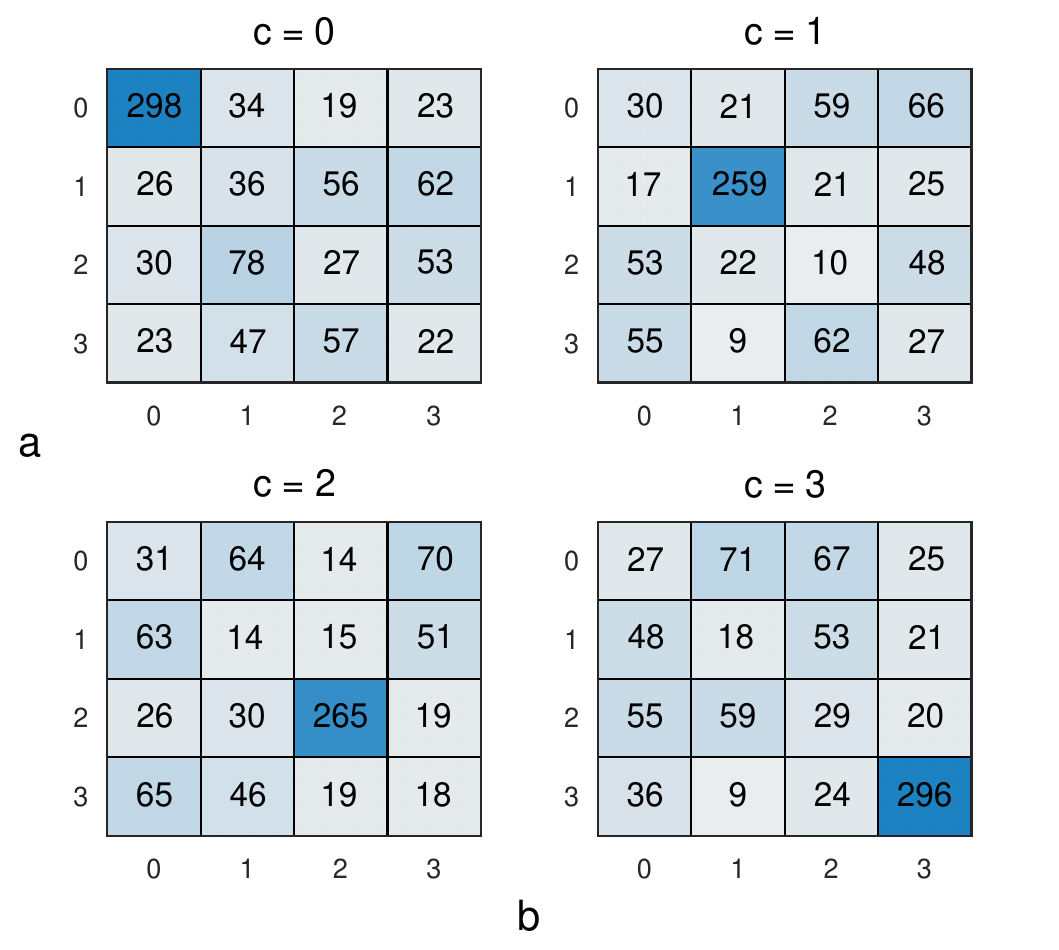}
    \caption{Six-fold coincidence counts corresponding to different measurement outputs. }
    \label{rawdata}
\end{figure}

\section{Comparison with previous studies}

In Ref.~\cite{baumer2021}, the authors use the IBM’s superconducting quantum computer to study the quantum network nonlocality which include the generation of the Elegant distribution in a triangle network. In order to show that they are close to the Elegant distribution, they calculate that the Kullback-Leibler (KL) divergence, $D(P||Q) = \sum_x P(x) \log \frac{P(x)}{Q(x)}$, between their distribution and $P_\text{E}$ is 0.272 (in contrast for us $D(P_{\text{Exp}}||P_{\text{E}}) = 0.0461\pm0.0052$.). However, this is insufficient to indicate nonlocality, as local distributions exist that are closer in KL divergence to the Elegent (e.g. the symmetric distribution with P(1,1,1) = 1/16 and P(1,2,3)=0.024 has a KL divergence of 0.044 to $P_\text{E}$). Moreover the amount of correlation expressed by $\sum_k P(k,k,k)$ is 0.281 for their distribution, less than the amount required (0.289) to violate the Bell inequalities which we significantly violate (see Supplamental Material Sec. I.C). 
Contrary to their results, our distribution is more correlated and closer to the Elegant distribution, allowing us to conclude its nonlocality with the portrayed tools.

    \clearpage
\end{widetext}

\end{document}